\documentclass[twocolumn,showpacs,preprintnumbers,amsmath,amssymb,superscriptaddress,prb]{revtex4-1}
\usepackage{graphicx}
\usepackage{dcolumn}
\usepackage{bm}

\begin{document}

\title{
  Spin wave modes in magnetic nanodisks under in-plane magnetic field 
}

\author{T. Kaneko}
\affiliation{Nano-scale theory group, NRI, AIST, 
  Tsukuba, Ibaraki, 305-8568, Japan}

\author{S. M. Noh}
\affiliation{Department of Electronic Engineering,
  Tohoku University, Sendai 980-8579 Japan}

\author{K. Miyake}
\affiliation{Department of Electronic Engineering,
  Tohoku University, Sendai 980-8579 Japan}

\author{M. Sahashi}
\affiliation{Department of Electronic Engineering,
  Tohoku University, Sendai 980-8579 Japan}

\author{H. Imamura}
\email{h-imamura@aist.go.jp}
\affiliation{Nano-scale theory group, NRI, AIST, 
  Tsukuba, Ibaraki, 305-8568, Japan}
\affiliation{Department of Electronic Engineering,
  Tohoku University, Sendai 980-8579 Japan}

\date{\today}

\begin{abstract}
 The size dependence of spin wave modes in a circular Permalloy (Py)
 nanodisk under an in-plane
 magnetic field is systematically studied by using micromagnetics simulations.
 We show that as the disk diameter is increased, the resonance frequency of
 the backward mode deceases while that of the
 uniform mode increases.  The avoided crossing of resonance frequencies
 of the uniform mode and the backward mode appears in the plot of the size
 dependence of resonance frequencies and the backward mode turns into
 the so-called ``edge mode'' for large nanodisks.
\end{abstract}


\maketitle

The dynamics of magnetizations confined in a magnetic nanostructure is of
special interest due to emerging applications in spintronics
devices, such as future recording head sensors, magnetic random-access
memories and spin torque oscillators. \cite{Lau_2011_JPhysD}
Ferromagnetic resonance (FMR) measurement is a powerful tool to
investigate the high-frequency response of magnetizations.  When 
magnetization is confined in a nanoscale structure, the FMR spectra of
the magnetic nanostructures show multi-peaks, i.e., several 
spin wave modes are excited. \cite{Walker1958jap,Noh2011ieeeTM} 

The spin wave modes of in-plane magnetized film are classified into
two modes depending on the relative direction of the magnetization
vector at equilibrium, $\bm{M}_{0}$, and the wavevector of the spin
wave, $\bm{q}$. \cite{Damon1960jap,KalinikosJPhysC1986}
One is the Damon and Eshbach (DE) mode where
$\bm{q}$ is perpendicular to $\bm{M}_{0}$.\cite{Damon1960jap}
The other is the backward (BA) mode where
$\bm{q}$ is parallel to $\bm{M}_{0}$.\cite{KalinikosJPhysC1986} 
These two modes have been observed in magnetic wires using Brillouin
light scattering(BLS). \cite{Jorzick1999prb,Mathieu1998prl}
For magnetic nanostructures, another spin wave mode called the ``edge 
mode'',  where the spin wave is localized around edges, appears. 
The edge mode has been observed in thin films,
\cite{Bailleul2001europhyslett,Jorzick2002prl} wires, circular disks
\cite{Gubbiotti2003prb,Giovannini2004prb,Neudecker2006,Shaw2009prb}
and elliptic disks. \cite{Gubbiottiprb2005,Jersch2010apl,Giovannini2011apl,Nembach2011prb}
It is important to study the relation among these spin wave modes in
magnetic nanostructures from both scientific and practical points of view. 

\begin{figure}[b!]
 \centering
 \includegraphics[width=\columnwidth]{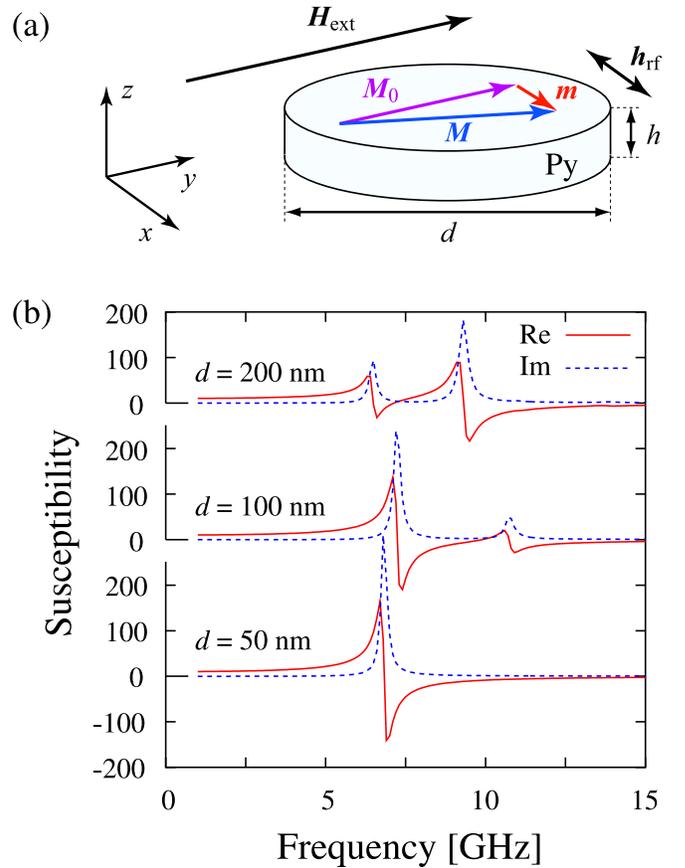}
 \caption{(Color online)
 (a) Schematic illustration of a circular Py nanodisk under static
 $\bm{H}_{\rm ext}$ and oscillating $\bm{h}_{\rm rf}$ magnetic fields. 
 (b) Real and imaginary parts of the calculated complex
 susceptibilities are plotted by the dotted and solid lines,
 respectively, for  $d$=50, 100 and 200 nm.
 }
 \label{fg:Fig1}
\end{figure}

In this paper, we systematically investigated the size dependence of
spin wave modes in the magnetic nanodisk shown in Fig.\ \ref{fg:Fig1} (a) by
using micromagnetics simulations. Until now it
has been widely accepted that the origin of the
edge mode is the pinning of the spin wave due to the strong
demagnetization field around the edges and that the edge mode has no
relation with the BA or DE modes. \cite{Bailleul2001europhyslett,Jorzick2002prl}
However, we show that the edge mode can be regarded as the BA mode
with two nodes.

\begin{figure*}[t!]
 \centering
 \includegraphics[width=2.0\columnwidth]{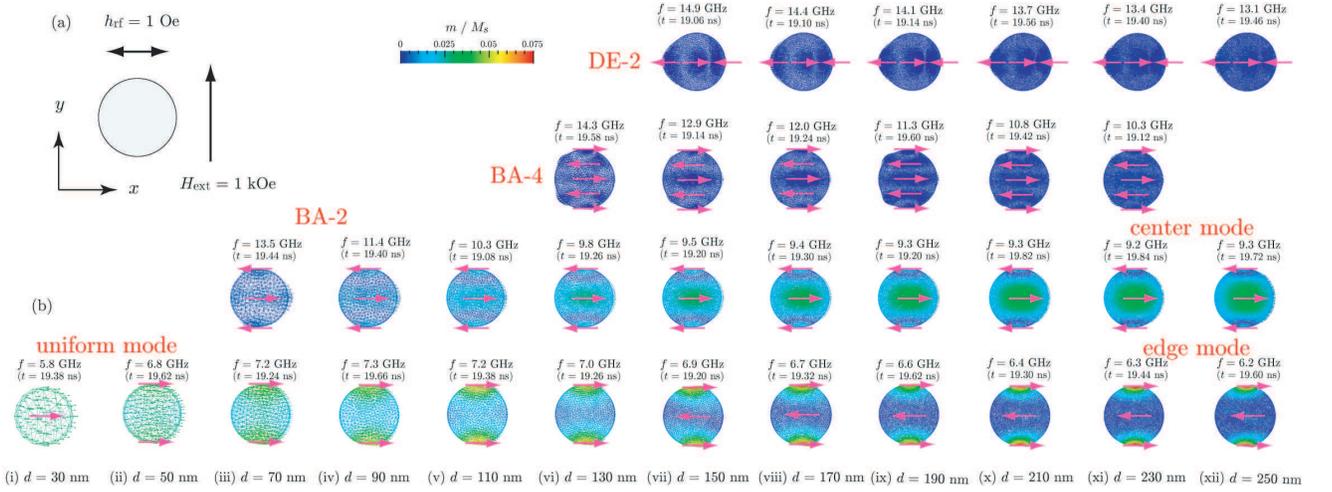}
 \caption{(Color online)
 (a) Directions of the static and oscillating magnetic fields are shown.
 (b) Snapshots of the distribution of the deviation,
 $\bm{m}(\bm{r},t)$, in a nanodisk with $d=30$ nm (i), 50 nm (ii), 70
 nm (iii), 90 nm (iv), 110 nm (v), 130 nm (vi), 150 nm (vii), 170 nm
 (viii), 190 nm (ix), 210 nm (x), 230 nm (xi) and 250 nm (xii). 
 The pink arrows indicate the directions of the deviation,
 $\bm{m}(\bm{r},t)$, at the anti-nodes.
 The labels BA-$n$ and DE-$n$ stand for the BA and DE modes with $n$
 nodal surfaces.
 }
  \label{fg:mode-all}
\end{figure*}

In order to solve the Landau-Lifshitz-Gilbert (LLG) equation for a
circular Py disk with thickness $h=10$ nm under an in-plane magnetic field we employed {\scriptsize NMAG} code
\cite{Fischbacher2007IEEE,NMAG} which is a micromagnetics simulation
package based on the finite-element and boundary-element methods.
We assumed a saturation magnetization of
$M_s=0.8\times 10^6$ A/m, an exchange stiffness is $A=1.3\times 10^{-11}$
J/m and a Gilbert dumping constant of $\alpha=0.01$.
The finite element mesh was generated using {\scriptsize Gmsh}. \cite{gmsh}
The size of the mesh was about 5 nm and we confirmed that this mesh was sufficiently fine
to describe the spin wave dynamics of the magnetic nanodisk we considered. 
The static magnetic field, $\bm{H}_{\rm ext}=H_{\rm ext}\bm{e}_y$, was
applied in the $y$-direction and the oscillating magnetic field,
$\bm{h}_{\rm rf} = h_{\rm rf}\sin(2\pi ft) \bm{e}_x$, of frequency $f$
was applied in the $x$-direction as shown in Fig.\ \ref{fg:Fig1} (a). 
We assumed $H_{\rm ext}=1.0$ kOe and $h_{\rm rf}=1$ Oe. 
We obtained a single domain state as an equilibrium magnetization
configuration for nanodisks with $d \le 250$ nm.  
Starting with the equilibrium magnetization configuration we
calculated the time evolution of the magnetization $\bm{M}(\bm{r},t)$
for 20 ns under the in-plane magnetic field of $\bm{H}_{\rm ext}+\bm{h}_{\rm rf}(t)$.

We introduce the deviation of the magnetization defined as
$\bm{m}(\bm{r},t)=\bm{M}(\bm{r},t)-\bm{M}_0(\bm{r})$, where
$\bm{M}_0(\bm{r})$ is the equilibrium magnetization.
The deviation $\bm{m}(\bm{r},t)$ represents the amplitude of a spin wave
excited by an oscillating magnetic field.
The real and imaginary part of the complex susceptibility $\chi$ was obtained 
by fitting the average of the $x$-component of the deviation, $\overline{m}_x$, to
\begin{equation}
  \overline{m}_x 
   = {\rm Re}(\chi) h_{\rm rf}\sin(2\pi ft)-{\rm Im}(\chi) h_{\rm rf}\cos(2\pi ft),
\end{equation}
after the oscillation was stabilized. In Fig.\ \ref{fg:Fig1} (b) we plot
the real and imaginary part of $\chi$ by the solid and dotted lines,
respectively for $d$=50, 100 and 200 nm.  The resonance frequencies 
were 6.8 GHz for $d=50$ nm, 7.2 and 10.7 GHz for $d=100$ nm and
6.5, 9.3, 11.0 and 13.8 GHz for $d=200$ nm. 
For  $d=200$ nm, the magnitudes of $\chi$ at 11.0 and 13.8 GHz were so small
that we used log-scale plots to identify them.
The calculated resonance frequencies were in good agreement
with those observed in TR-MOKE measurement of circular Py nanodisks with $h=$10
nm by Shaw {\it et al}. They observed about 7 GHz for
$d=50$ nm, 7 and 10 GHz, for $d=100$ nm, and 7 and 9 GHz for $d=200$ nm under
the in-plane magnetic field of 1 kOe \cite{Shaw2009prb}.

Figure \ref{fg:mode-all} (b) shows snapshots of the deviation,
$\bm{m}(\bm{r},t)$, at the resonance frequencies for $d=30,50,\cdots,250$
nm. The directions of the static and oscillating magnetic fields in this
plot are shown in Fig.\ \ref{fg:mode-all} (a). 
The labels BA-$n$ and DE-$n$ stand for the BA and DE modes with $n$
nodal surfaces as shown in Fig.\ \ref{fg:mode-all} (b).
The pink arrows indicate the directions of the deviation,
$\bm{m}(\bm{r},t)$, at the anti-nodes.
We could not identify the third lowest mode for $d=250$ nm since its resonance
frequency was very close to that of the center mode. 
For $d=30$ nm, the deviation, $\bm{m}$, was uniformly distributed throughout
the disk, i.e., the uniform mode was excited. \cite{Kittel1948pr}
As the disk diameter, $d$, is increased, other resonant spin wave modes
classified as BA-2, BA-4 and DE-2 modes appear.
It should be noted that for a large nanodisk the lowest mode is not the
uniform mode but the edge mode, and the second
lowest mode is not the BA-2 mode but the center mode.
Tracing the spin wave modes of the lowest-resonance frequencies, i.e.,
the bottom line of Fig.\ \ref{fg:mode-all} (b), one might
come up with the idea that the origin of the edge mode is the uniform
mode and it has no relation with either the BA or DE modes.
However, our systematic investigation of the resonant spin wave modes
revealed that the above interpretation is not valid.

\begin{figure}[t!]
  \centering
  \includegraphics[width=\columnwidth]{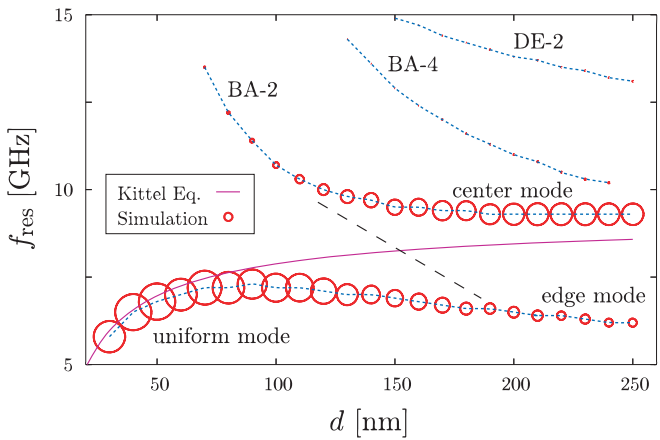}
  \caption{(Color online)
 Calculated resonance frequencies,$f_{\rm res}$, are plotted by circles
 with radii proportional to imaginary part of the susceptibility,
 ${\rm Im}(\chi)$, against the disk diameter, $d$.
 The blue dotted lines are visual guides connecting resonance
 frequencies in the same branches. 
 The purple solid line represents the resonance frequencies given by 
 the Kittel equation of  Eq. (\protect\ref{eq:KittelEq}). The labels of the
 excited spin wave modes are the same as those in Fig. \protect\ref{fg:mode-all} (b).
  }
  \label{fg:d-f_res}
\end{figure}

In Fig.\ \ref{fg:d-f_res} we plot the size dependence of the
calculated resonance frequencies by circles with radii proportional to
the imaginary part of the susceptibility, ${\rm Im}(\chi)$.
The blue dotted lines are visual guides connecting resonance
frequencies in the same branches. 
The purple solid line represents the resonance frequencies given by 
the Kittel equation \cite{Kittel1948pr}:
\begin{equation}
  f_{\rm K} = \sqrt{f_H(f_H+(N_z-N_y)f_M)},
  \label{eq:KittelEq}
\end{equation}
where $f_H=\gamma H_{\rm ext}/2\pi$, $f_M=\gamma M_s/2\pi$, $\gamma$
is the gyromagnetic ratio and $N$'s represent the demagnetization factors
estimated by means of the micromagnetics simulations.
For small nanodisks with $d \le 60$ nm, we obtained a single resonance peak
at frequencies close to those given by the Kittel equation of
Eq.\ \eqref{eq:KittelEq}.
As the disk diameter, $d$, is increased, the frequency of the uniform mode
increases due to the increase of the demagnetization field, while those of
the BA-2, BA-4 and DE-2 modes decrease. At around $d\simeq 150$ nm where
the frequency of the BA-2 mode comes close to
that of the uniform mode, an avoided crossing of the BA-2 mode and uniform mode
appears and the BA-2 mode becomes the edge mode for large disks as
indicated by the black dashed line. 
Therefore, the edge mode can be regarded as the BA-2 mode with two nodes.
The uniform mode becomes the center mode, the frequency of which is close
to that given by the Kittel equation.

In summary, we investigated the size dependence of spin wave modes in
a circular Py nanodisk with $d \le$ 250 nm under an in-plane magnetic field
using micromagnetics simulations.
We showed that the excited spin wave modes are classified into uniform
mode, edge mode, center mode, BA-2 mode, BA-4 mode, and DE-2 mode
depending on the size of the nanodisk.
For large disks, we found an avoided crossing of the BA-2 and uniform modes,
at which BA-2 mode (uniform mode) becomes the edge mode (center mode).

The authors acknowledge M.\ Doi and H.\ Arai for valuable discussions.
This work was supported in part through NEDO, the Storage Research
Consortium (SRC) and Center of Education, Research for Information
Electronics Systems of Global COE Program, and MEXT.

\end{document}